\documentclass[reqno]{amsart}

\usepackage{amssymb,epsf}

\def\be{\begin{equation}}
\def\ee{\end{equation}}
\def\ba{\begin{align}}
\def\ea{\end{align}}
\def\bsplit{\begin{split}}
\def\esplit{\end{split}}
\def\bm{\begin{multline}}
\def\eem{\end{mutline}}
\def\bfig{\begin{figure}[htb]}
\def\efig{\end{figure}}

\numberwithin{equation}{section}
\newtheorem{theorem}{THEOREM}

\newcommand{\dd}{{\rm d}}
\newcommand{\e}[1]{\,{\rm e}^{#1}\,}
\newcommand{\ii}{{\rm i}}
\newcommand{\sumtwo}[2]{\sum_{\substack{#1 \\ #2}}}

\DeclareMathOperator*{\supp}{\text{supp}}

\def\Tr{{\operatorname{Tr\,}}}

\newcommand{\upchi}{\raise 2pt \hbox{$\chi$}}

\newcommand{\caG}{{\mathcal G}}

\newcommand{\caS}{{\mathcal S}}

\newcommand{\bbN}{{\mathbb N}}

\newcommand{\bbR}{{\mathbb R}}

\newcommand{\bbZ}{{\mathbb Z}}

\newcommand{\bsx}{{\boldsymbol x}}

\newcommand{\bsvarrho}{{\boldsymbol\varrho}}

\begin{document}


\title[Interacting spatial permutations and Bose gas]{The model of interacting spatial permutations and its relation to the Bose gas}

\author{Daniel Ueltschi}
\address{Department of Mathematics \hfill\newline
\indent University of Warwick \hfill\newline
\indent Coventry, CV4 7AL, England \hfill\newline
{\small\rm\indent http://www.ueltschi.org}
}
\email{daniel@ueltschi.org}

\begin{abstract}
The model of spatial permutations is related to the Feynman-Kac representation of the Bose gas.
The transition to infinite cycles corresponds to Bose-Einstein condensation.
We review the general setting and some results, and we derive a multi-body interaction between permutation jumps, that is due to the original interactions between quantum particles.

\vspace{1mm}

\vspace{1mm}
\noindent
{\it 2000 Math.\ Subj.\ Class.:} 60K35, 82B20, 82B26, 82B41
\end{abstract}

\thanks{\copyright\, 2007 by the author. This article can be reproduced, in its entirety, for non-commercial purposes.}

\maketitle

\section{Introduction}

One purpose of this article is to review the setting for the model of {\it spatial permutations} and its relation with the quantum Bose gas, and to summarize some of the material presented in a recent collaboration with Volker Betz \cite{BU}.
Another purpose is to compute the effective interaction between permutation jumps.
It involves the original interaction potential between quantum particles.
While several mathematical questions remain unanswered, it is argued that the model of interacting spatial permutations describes the quantum interacting Bose gas {\it exactly}, and in a simpler way.
The main phenomenon in bosonic systems is the {\it Bose-Einstein condensation}.
We discuss the links between this phase transition and the occurrence of {\it infinite cycles} in random permutations.

Given points $x_1,\dots,x_N$ in $\bbR^d$, one considers random permutations $\pi$ of $N$ elements with weight
\[
\prod_{i=1}^N \exp \bigl\{ -\tfrac1{4\beta} |x_i - x_{\pi(i)}|^2 \bigr\}.
\]
Permutation jumps are essentially finite, but permutation cycles can be large.
This model is illustrated in Fig.\ \ref{figpermutation}.
It is motivated in large part by the Feynman-Kac representation of the Bose gas.
We actually discuss a more general setting where permutation jumps interact.

\bfig
\epsfxsize=80mm \centerline{\epsffile{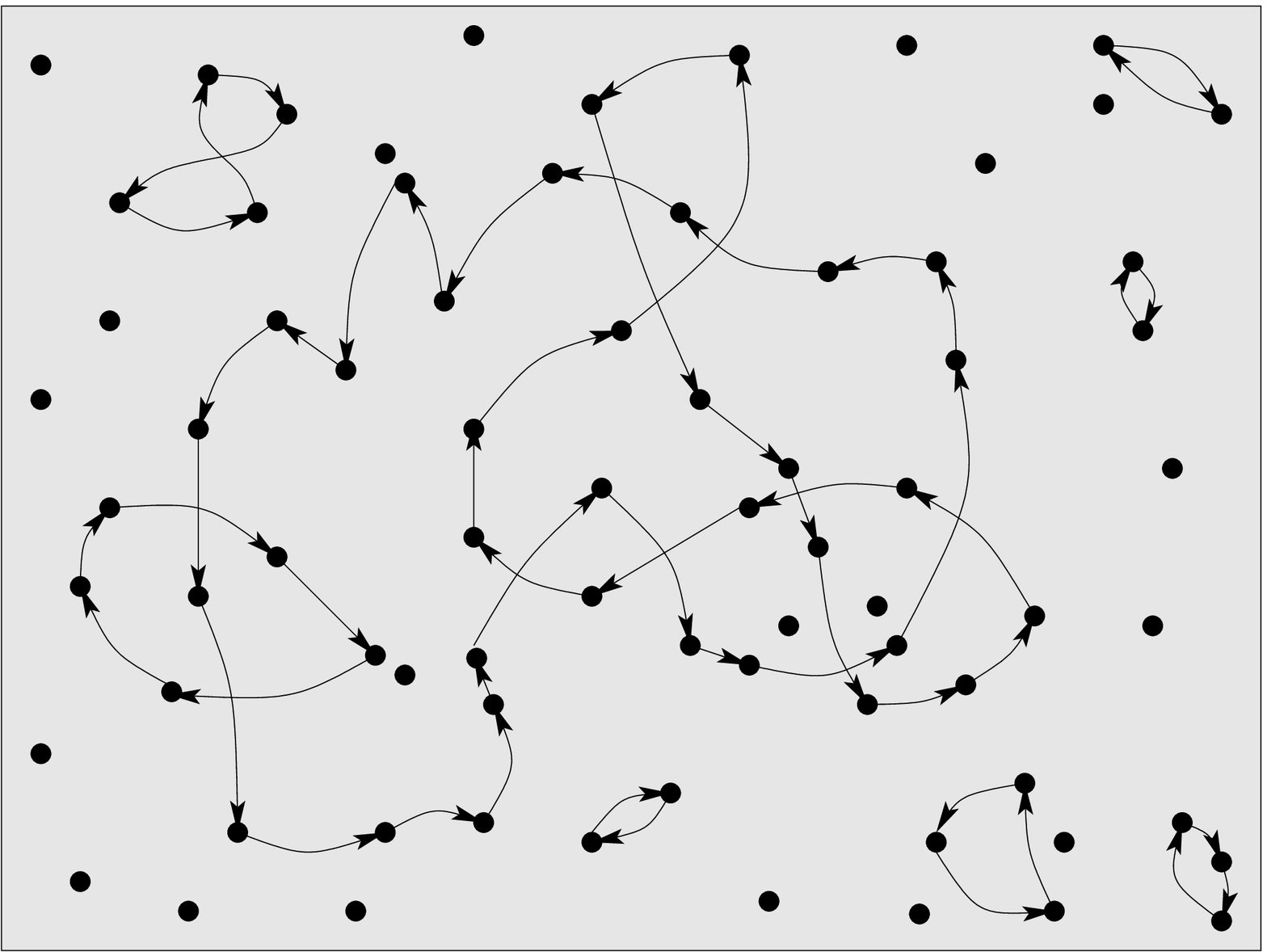}}
\caption{Illustration for a random set of points $\bsx = (x_1,\dots,x_N)$, and for a permutation $\pi \in \caS_N$.
Isolated points are sent onto themselves.
Permutation jumps are small, but long cycles can occur nonetheless.}
\label{figpermutation}
\efig

The precise setting is introduced in Section \ref{secsetting}.
We recall the Feynman-Kac representation of the Bose gas in Section \ref{secFK};
it makes the relation between the ideal Bose gas and non-interacting spatial permutations clear.
The two-body interaction between permutation jumps, that is expected to give the exact behaviour to lowest order in the strength of the particle interactions, is computed in Section \ref{secinteractions}.
Finally, we describe a simple model of interacting permutations in Section \ref{sectoymodel}.
It is exactly solvable, and it provides some understanding about the effects of interactions on the Bose-Einstein condensation.

\section{The model of spatial permutations}
\label{secsetting}

Let $\Lambda \subset \bbR^d$ be a cube of size $L$ and volume $V = L^d$, and let $N \in \bbN$.
The state space of the model of spatial permutations is
\be
\Omega_{\Lambda,N} = \Lambda^N \times \caS_N,
\ee
with $\caS_N$ the symmetric group of permutations of $N$ elements.
We are interested in the properties of permutations, and all our random variables are functions $\theta : \caS_N \to \bbR$.
Their probability distributions depend on spatial variables in an indirect but essential way.
Let $\ell_i(\pi)$ denote the length of the cycle that contains $i$, i.e.\ the smallest integer $n\geq1$ such that $\pi^n(i) = i$.
The most important random variable is the density of points in cycles of certain lengths.
For $n,n' \in \bbN$, let
\be
\label{defdenscycles}
\bsvarrho_{n,n'}(\pi) = \frac1V \, \#\bigl\{ i=1,\dots,N : n \leq \ell_i(\pi) \leq n' \bigr\}.
\ee

The expectation of the random variable $\theta$ is defined by
\be
\label{def expectation}
E_{\Lambda,N}(\theta) = \frac1{Z(\Lambda,N) N!} \int_{\Lambda^N} \dd \bsx \sum_{\pi\in\caS_N} \theta(\pi) \e{-H(\bsx,\pi)}.
\ee
Here, the normalization factor $Z(\Lambda,N)$ is chosen so that $E_{\Lambda,N}(1) = 1$.
The term $N!$ is present in order that $Z(\Lambda,N)$ scales like the exponential of the volume of $\Lambda$ --- thus behaving like a partition function in statistical mechanics.
The integral is over $N$ points in $\Lambda$, denoted $\bsx = (x_1,\dots,x_N)$.

We consider Hamiltonians of the form
\be
\label{def Ham}
H(\bsx,\pi) = \sum_{i=1}^N \xi(x_i - x_{\pi(i)}) + \sum_{1\leq i<j\leq N} V(x_i, x_{\pi(i)}, x_j, x_{\pi(j)}),
\ee
with $\xi$ a spherically symmetric function $\bbR^d \to [0,\infty]$, and $V$ a translation invariant function $\bbR^{4d} \to \bbR$.
We also suppose that $\xi$ is increasing and that $\xi(0)=0$.
One should think of typical permutations as involving finite jumps, i.e.\ $|x_i - x_{\pi(i)}|$ stays bounded as $\Lambda,N \to \infty$.

The major question concerns the occurrence of infinite cycles.
It turns out that the distribution of cycles can be well characterized in the absence of interactions, with the potential $V \equiv0$.
We need a few hypotheses on $\xi$.
Let $C = \int \e{-\xi}$.
We suppose that $\e{-\xi}$ has positive Fourier transform, which we denote $C \e{-\varepsilon(k)}$.
Precisely, we have
\be
C \e{-\varepsilon(k)} = \int_{\bbR^d} \e{-2\pi\ii kx} \e{-\xi(x)} \dd x.
\ee
The case of physical relevance is $\xi(x) = \frac1{4\beta} |x|^2$ with $\beta$ the inverse temperature, in which case $\varepsilon(k) = 4\pi^2 \beta |k|^2$.
But it may be of mathematical interest to consider other functions, including some where $\e{-\xi}$ has bounded support.
Criteria that guarantee positivity of the Fourier transform are discussed e.g.\ in \cite{HS}.

We define the {\it critical density} by
\be
\label{critdens}
\rho_{\rm c} = \int_{\bbR^d} \frac{\dd k}{\e{\varepsilon(k)} - 1}.
\ee
The critical density is finite for $d\geq3$, but it can be infinite for $d=1,2$.
The experienced physicist will have recognized the formula for the critical density of Bose-Einstein condensation.
The relation with the Bose gas will be discussed in the next section.
In the following theorem we fix the density $\rho$ and we let $N = \rho V$ in the expectation \eqref{def expectation}.

\begin{theorem}
\label{thminfinitecycles}
Let $\xi$ satisfy the assumptions above. Then for any $0<a<b<1$, and any $s\geq0$,
\[
\begin{split}
&{\rm (a)} \quad \lim_{V\to\infty} E_{\Lambda, \rho V}(\bsvarrho_{1,V^a}) = \begin{cases} \rho & \text{if } \rho \leq \rho_{\rm c}; \\ \rho_{\rm c} & \text{if } \rho \geq \rho_{\rm c}; \end{cases} \\
&{\rm (b)} \quad \lim_{V\to\infty} E_{\Lambda, \rho V}(\bsvarrho_{V^a,V^b}) = 0; \\
&{\rm (c)} \quad \lim_{V\to\infty} E_{\Lambda, \rho V}(\bsvarrho_{V^b,sV}) = \begin{cases} 0 & \text{if } \rho \leq \rho_{\rm c}; \\ s & \text{if } 0 \leq s \leq \rho-\rho_{\rm c}; \\
\rho - \rho_{\rm c} & \text{if } 0 \leq \rho-\rho_{\rm c} \leq s. \end{cases}
\end{split}
\]
\end{theorem}

In order to understand the meaning of these claims, one should think of $a$ as barely bigger than 0, and $b$ barely smaller than 1.
In part (a), $\bsvarrho_{1,V^a}$ is the density of points in finite cycles.
All points belong to finite cycles if $\rho \leq \rho_{\rm c}$.
However, if $\rho > \rho_{\rm c}$, a fraction $\rho-\rho_{\rm c}$ of points belong to infinite cycles.
It is natural to ask oneselves about the size of ``infinite cycles'' in a finite domain of volume $L^d=V$.
One could expect the typical length to be of order $L^2$, since the continuum limit of random walks has Hausdorff dimension 2, and cycles are somewhat like closed random walks.
However, part (b) shows that cycles of length $V^b$, with $0<b<1$, have vanishing density.
Thus infinite cycles are {\it macroscopic}, i.e.\ each cycle involves a strictly positive fraction of points.
The statistics of macroscopic cycles is characterized in part (c).

The proof of Theorem \ref{thminfinitecycles} can be found in \cite{BU}.
Actually, the correct statement involves periodic boundary conditions; the interested reader is invited to look in \cite{BU} for the precise statement.
Theorem \ref{thminfinitecycles} extends an earlier result of S\"ut\H o for the ideal Bose gas \cite{Suto2}.

A different model is investigated in \cite{GRU}, where the positions $x_1,\dots,x_N$ form a cubic lattice.
The density is always equal to 1, but the function $\xi(x)$ depends on a parameter that represents the temperature of the system.
It is found numerically that the critical temperature for the occurrence of infinite cycles is close but different from that of the ideal Bose gas.
Many properties are similar, however; infinite cycles are also macroscopic.
A surprising fact is that the expectation of the length of the longest cycle seems to be identical to that in the ideal Bose gas;
this suggests that the distribution of macroscopic cycles may be the same.

\section{Feynman-Kac representation of the Bose gas}
\label{secFK}

The Feynman-Kac formula relates the kernel of $\e{-\beta H}$, with $H$ a Schr\"odinger operator, to the Brownian motion, whose mathematical expression is the Wiener measure.
It seems to have first appeared in \cite{Fey}, precisely in the context of bosonic systems and in the discussion of cycles.
Ginibre wrote an excellent mathematical introduction to the Wiener measure, the Feynman-Kac formula, and its application to bosonic systems \cite{Gin}.
We review these notions here without introducing the full mathematical setting, but all equations below can be justified with a bit of analysis.
In particular, we do not discuss the details arising from the boundary conditions;
as usual in statistical mechanics, they are irrelevant for large systems.

Let $\Lambda \subset \bbR^d$ be a cube of size $L$, and let $g_\beta$ denote the normalized Gaussian function
\be
g_\beta(x) = \frac1{(2\pi\beta)^{d/2}} \e{-|x|^2 / 2\beta}.
\ee
It is not hard to check that
\be
\int_\Lambda g_s(x-a) g_t(x-b) \dd x = g_{s+t}(a-b),
\ee
and that, after iteration,
\be
\label{produit de gaussiennes}
\int_{\Lambda^{n-1}} \dd x_2 \dots \dd x_n \prod_{i=1}^n g_t(x_{i+1}-x_i-a_i) = g_{nt} \bigl( x_{n+1} - x_1 - \Sigma_{i=1}^n a_i \bigr).
\ee

Let $H = -\Delta + U$ be a Schr\"odinger operator in $L^2(\Lambda)$, with $\Delta$ the Laplacian and $U$ a smooth real function.
This operator is unbounded and we need to specify its domain.
We can choose the space of $C^2$ functions on $\Lambda$ with Dirichlet boundary conditions.
Then $H$ is symmetric and we consider its self-adjoint extension.
Of relevance to statistical mechanics is the operator $\e{-\beta H}$.
It is a nice operator, bounded and compact, but these properties are not important here.
The Feynman-Kac formula states that (with $x_{n+1} \equiv x_1$)
\be
\bsplit
\Tr \e{-\beta H} &= \lim_{n\to\infty} \int_{\Lambda^n} \dd x_1 \dots \dd x_n \Bigl[ \prod_{i=1}^n g_{2\beta/n}(x_{i+1}-x_i) \Bigr] \exp \Bigl\{ -\frac\beta n \sum_{i=1}^n U(x_i) \Bigr\} \\
&\equiv \int_\Lambda \dd x \, \exp \Bigl\{ -\tfrac12 \int_0^{2\beta} U \bigl( \omega(s) \bigr) \dd s \Bigr\} \dd W_{xx}^{2\beta}(\omega).
\end{split}
\ee
Here, $\omega$ is a Brownian bridge starting and ending at $x$ and traveling in time $2\beta$, and $W_{xx}^{2\beta}$ is the Wiener measure.
In the second line we should restrict the paths to stay inside $\Lambda$, because of Dirichlet boundary conditions;
we neglect these technicalities, however.

Let us turn to the description of bosonic systems.
The state space for $N$ quantum bosons in a domain $\Lambda \subset \bbR^d$ is the subspace $L^2_{\rm sym}(\Lambda^N)$ of symmetric complex functions of $N$ variables.
The Hamiltonian is given by the Schr\"odinger operator
\be
H = -\sum_{i=1}^N \Delta_i + \sum_{1 \leq i,j \leq N} U(x_i - x_j).
\ee
Here, $\Delta_i$ denotes the $d$-dimensional Laplacian for the $i$-th variable, and $U(x_i - x_j)$ is a multiplication operator that represents the interaction between particles $i$ and $j$.
We always suppose that $U(x) \geq 0$.
We can choose the self-adjoint extension of $H$ that corresponds to Dirichlet boundary conditions.
Of course, the sum of $N$ Laplacians in $\Lambda^d$ can be viewed as a Laplacian in $\Lambda^{dN}$, so we can apply the Feynman-Kac formula.

The canonical partition function is equal to
\be
\Tr_{L^2_{\rm sym}(\Lambda^N)} \e{-\beta H} = \Tr_{L^2(\Lambda^N)} P_+ \e{-\beta H}
\ee
where $P_+$ is the projector onto symmetric functions,
\be
P_+ \varphi(x_1,\dots,x_N) = \frac1{N!} \sum_{\pi\in\caS_N} \varphi(x_{\pi(1)},\dots,x_{\pi(N)}).
\ee

Using this projection and the Feynman-Kac formula, the partition function of the Bose gas can be written as
\be
\Tr \e{-\beta H} = \frac1{N!} \int_{\Lambda^N} \dd\bsx \sum_{\pi\in\caS_N} \e{-H'(\bsx,\pi)},
\ee
with the Gibbs factor given by
\be
\label{Gibbs for Bose}
\e{-H'(\bsx,\pi)} = \Bigl[ \prod_{i=1}^N \int\dd W_{x_i x_{\pi(i)}}^{2\beta}(\omega_i) \Bigr] \exp \Bigl\{ -\tfrac12 \sum_{1\leq i<j\leq N} \int_0^{2\beta} U \bigl( \omega_i(s) - \omega_j(s) \bigr) \dd s \Bigr\}.
\ee
This formula is illustrated in Fig.\ \ref{figfeykac}.
It involves spatial positions and permutations of these positions.
\bfig
\epsfxsize=80mm \centerline{\epsffile{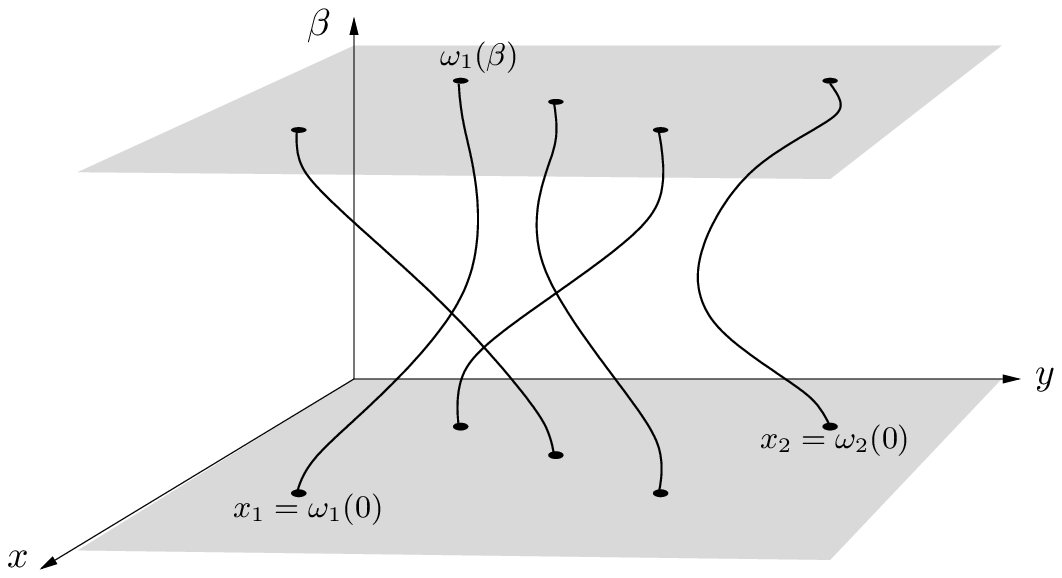}}
\caption{Feynman-Kac representation of a gas of $N$ bosons.
The horizontal plane represents the $d$ spatial dimensions, and the vertical axis is the imaginary time dimension.
The picture shows five particles and two cycles, of respective length 4 and 1.}
\label{figfeykac}
\efig

In the case of the ideal gas, $U \equiv 0$, the paths no longer interact and the Wiener integrals in \eqref{Gibbs for Bose} can be computed.
We find that
\be
\label{ideal Gibbs}
\e{-H'(\bsx,\pi)} = (4\pi\beta)^{-dN/2} \e{-H^{(0)}(\bsx,\pi)}
\ee
with
\be
\label{def H0}
H^{(0)}(\bsx,\pi) = \frac1{4\beta} \sum_{i=1}^N \bigl| x_i - x_{\pi(i)} \bigr|^2.
\ee
The prefactor in \eqref{ideal Gibbs} plays no r\^ole in expectations of random variables and it can be ignored.
Thus the ideal Bose gas is equivalent to the ``ideal'' model of spatial permutations with Gaussian weights.
Random variables of permutations have same distribution in both models, and the transition to infinite cycles takes place at the same critical density.

The equivalence between the occurrence of infinite cycles and Bose-Einstein condensation is an open problem.
It is known to be true in the ideal gas, see \cite{Suto1,Suto2,Uel1}, but it does not seem to be true in strongly interacting systems in a solid phase.
Pollock and Ceperley have argued that superfluidity is related to {\it spatially winding} cycles \cite{PC}.
Such cycles are clearly infinite in the thermodynamic limit.
On the other hand, we know from Theorem \ref{thminfinitecycles} that infinite cycles are macroscopic (i.e.\ they have strictly positive density), so they certainly have non-zero winding number.
Infinite and winding cycles should therefore be equivalent.
Superfluidity is by no means equivalent to Bose-Einstein condensation.
These facts bring some level of confusion and we can only hope that they will be clarified in the future.

However, it is expected that, in dimension $d\geq3$, weakly interacting bosonic systems have the {\it same critical density} for Bose-Einstein condensation, superfluidity, infinite cycles, and winding cycles.
Hereafter, we study the occurrence of infinite cycles in the weakly interacting regime, and we implicitly assume that they reveal a Bose-Einstein condensation.

\section{Exact two-body interaction for permutation jumps}
\label{secinteractions}

\subsection{Expansion of path interactions}

The two-body interactions between quantum particles translate into many-body interactions for permutations.
But we can perform an expansion and see that, to lowest order, we obtain a two-body interaction between permutation jumps.

Let $\widehat W^t_{x,y} = g_t^{-1}(x-y) W_{x,y}^t$ be a Wiener measure normalized such that $\int\dd\widehat W^t_{x,y}(\omega) = 1$.
From \eqref{Gibbs for Bose}, we have
\be
\e{-H'(\bsx,\pi)} = (4\pi\beta)^{-dN/2} \e{-H^{(0)}(\bsx,\pi)} \e{-H^{(1)}(\bsx,\pi)}
\ee
with $H^{(0)}$ given by \eqref{def H0}, and
\be
\label{ca commence}
\begin{split}
\e{-H^{(1)}(\bsx,\pi)} &= \Bigl[ \prod_{i=1}^N \int\dd\widehat W_{x_i x_{\pi(i)}}^{2\beta}(\omega_i) \Bigr] \prod_{1\leq i<j\leq N} \e{-\frac12 \int_0^{2\beta} U \bigl( \omega_i(s) - \omega_j(s) \bigr) \dd s} \\
&= \Bigl[ \prod_{i=1}^N \int\dd\widehat W_{x_i x_{\pi(i)}}^{2\beta}(\omega_i) \Bigr] \prod_b \bigl( 1 - \Upsilon(\omega_b) \bigr).
\end{split}
\ee
The last product is over bonds $b = \{i,j\}$ with $i \neq j$.
For $i<j$, we defined $\omega_b = \omega_i - \omega_j$, and
\be
\Upsilon(\omega_b) = 1 - \e{-\frac12 \int_0^{2\beta} U(\omega_b(s)) \dd s}.
\ee
Expanding the product in \eqref{ca commence}, we have
\be
\label{developpement}
{}\e{-H^{(1)}(\bsx,\pi)} = \Bigl[ \prod_{i=1}^N \int\dd\widehat W_{x_i x_{\pi(i)}}^{2\beta}(\omega_i) \Bigr] \sum_{k=0}^{\frac12 N(N-1)} (-1)^k \sum_{\{b_1,\dots,b_k\}} \prod_{m=1}^k \Upsilon(\omega_{b_m}).
\ee
In the regime of weak interactions, the typical $k$ in the above sum is a small fraction of the volume, and the typical $b_1,\dots,b_k$ are mostly disjoint.

We first perform the expansion in a somewhat cavalier fashion.
We will be more precise in Section \ref{secclexp}, where we will check that we have identified the leading order.
Let $b_m = \{i_m,j_m\}$.
Let us assume that $b_\ell \cap b_m = \emptyset$ for all $\ell \neq m$; then
\be
\label{identite partielle}
\Bigl[ \prod_{i=1}^N \int\dd\widehat W_{x_i x_{\pi(i)}}^{2\beta}(\omega_i) \Bigr] \prod_{m=1}^k \Upsilon(\omega_{b_m}) = \prod_{m=1}^k V(x_{i_m}, x_{\pi(i_m)}, x_{j_m}, x_{\pi(j_m)})
\ee
where the potential $V$ has been defined by
\be
\label{def 2 body interaction}
V(x,y,x',y') = \int\dd\widehat W_{xy}^{2\beta}(\omega) \int\dd\widehat W_{x'y'}^{2\beta}(\omega') \Upsilon(\omega-\omega').
\ee
This is the two-body interaction between jumps $x \mapsto y$ and $x' \mapsto y'$.
The expression \eqref{def 2 body interaction} can be simplified, see Eq.\ \eqref{LaFormule} below.
We use the identity \eqref{identite partielle} for all $b_1,\dots,b_k$ that appear in \eqref{developpement}, not only disjoint ones.
This is an approximation;
it assumes that either the terms with intersecting $b_m$'s are not important, or that their contribution is close to \eqref{identite partielle}.
We obtain
\be
{}\e{-H^{(1)}(\bsx,\pi)} \approx \sum_{k\geq0} (-1)^k \sum_{\{b_1,\dots,b_k\}} \prod_{m=1}^k V(x_{i_m}, x_{\pi(i_m)}, x_{j_m}, x_{\pi(j_m)}).
\ee
Ignoring the possibility that a same bond may occur several times, we get
\be
\label{approximation}
\begin{split}
\e{-H^{(1)}(\bsx,\pi)} &\approx \sum_{k\geq0} \frac{(-1)^k}{k!} \sum_{b_1,\dots,b_k} \prod_{m=1}^k V(x_{i_m}, x_{\pi(i_m)}, x_{j_m}, x_{\pi(j_m)}) \\
&= \exp \Bigl\{ -\sum_{1\leq i<j\leq N} V(x_i, x_{\pi(i)}, x_j, x_{\pi(j)}) \Bigr\}.
\end{split}
\ee
These approximations suggest that, to lowest order in the strength of the interaction, the multi-body interaction arising from the Feynman-Kac representation can be approximated by the two-body interaction defined in \eqref{def 2 body interaction}.

\subsection{Cluster expansion}
\label{secclexp}

It is not clear that the approximations above have produced the correct terms, that are exact to lowest order.
In this section we perform a cluster expansion.
It cannot be entirely justified from a mathematical point of view, but it nevertheless clarifies the approximations.

Consider the graph with vertices $\{1,\dots,k\}$, and with an edge between $\ell$ and $m$ whenever $b_\ell \cap b_m \neq \emptyset$.
We say that a set of bonds $B = \{b_1,\dots,b_k\}$ is {\it connected} if this graph is connected.
Let $\supp B = \cup_{b\in B} b$.
We say that $B$ and $B'$ are {\it compatible} if their supports are disjoint, $\supp B \cap \supp B' = \emptyset$.
Then the sum over sets of bonds in \eqref{developpement} can be written as a sum over connected and mutually compatible $B$'s, namely
\be
\sum_{k\geq0} (-1)^k \sum_{\{b_1,\dots,b_k\}} \prod_{m=1}^k \Upsilon(\omega_{b_m}) = \sum_{\ell\geq0} \frac1{\ell!} \sumtwo{B_1,\dots,B_\ell}{\text{compatible}} \prod_{m=1}^\ell \Bigl[ (-1)^{|B_m|} \prod_{b \in B_m} \Upsilon(\omega_b) \Bigr].
\ee
The contribution of compatible $B$'s factorizes.
For a connected $B$, let us introduce
\be
\Phi_B(\bsx,\pi) = (-1)^{|B|} \Bigl[ \prod_{i \in \supp B} \int\dd\widehat W_{x_i x_{\pi(i)}}^{2\beta}(\omega_i) \Bigr] \prod_{b\in B} \Upsilon(\omega_b).
\ee
Notice that $\Phi_B(\bsx,\pi)$ depends only on positions $x_i$ and $x_{\pi(i)}$ for $i \in \supp B$.
Then we have
\be
\label{encore un developpement}
\e{-H^{(1)}(\bsx,\pi)} = \sum_{\ell\geq0} \frac1{\ell!} \sumtwo{B_1,\dots,B_\ell}{\text{compatible}} \prod_{m=1}^\ell \Phi_{B_m}(\bsx,\pi).
\ee

We now apply the cluster expansion method, see e.g.\ \cite{KP,Uel} for references.
Given $B_1,\dots,B_\ell$, let $\varphi(B_1,\dots,B_\ell)$ be the following combinatorial function:
\be
\varphi(B_1,\dots,B_\ell) = \begin{cases} 1 & \text{if } \ell=1; \\ \displaystyle \frac1{\ell!} \sum_{G \subset \caG(B_1,\dots,B_\ell)} (-1)^{|G|} & \text{if } \ell \geq 2. \end{cases}
\ee
Here, $\caG(B_1,\dots,B_\ell)$ denotes the graph with $\ell$ vertices, and with an edge between $i$ and $j$ whenever $B_i$ and $B_j$ are {\it not} compatible.
The sum is over all {\it connected} subgraphs of $\ell$ vertices, and $|G|$ is the number of edges of $G$.
Notice that $\varphi(B_1,\dots,B_\ell)$ is zero unless $B_1,\dots,B_\ell$ form a cluster, i.e.\ unless $\caG(B_1,\dots,B_\ell)$ is connected.

The cluster expansion yields a convergent series for the logarithm of \eqref{encore un developpement}, hence for $H^{(1)}$.
Precisely,
\be
\label{clexp}
H^{(1)}(\bsx,\pi) = -\sum_{\ell\geq1} \sum_{B_1,\dots,B_\ell} \varphi(B_1,\dots,B_\ell) \prod_{m=1}^\ell \Phi_{B_m}(\bsx,\pi).
\ee
Let $i_1,\dots,i_k$ be distinct indices.
The previous equation suggests to define the $k$-body interaction by
\be
\label{def Vk}
V^{(k)} \bigl( (x_{i_\ell},x_{\pi(i_\ell)})_{\ell=1}^k \bigr) = -\sum_{m\geq1} \sumtwo{B_1,\dots,B_m}{\cup_\ell \supp B_\ell = \{i_1,\dots,i_k\}} \varphi(B_1,\dots,B_m) \prod_{\ell=1}^m \Phi_{B_\ell}(\bsx,\pi).
\ee
Then $H^{(1)}$ is given by
\be
\label{toutes les interactions}
H^{(1)}(\bsx,\pi) = \sum_{k\geq2} \; \sum_{1 \leq i_1 < \dots < i_k \leq N} V^{(k)} \bigl( (x_{i_\ell}, x_{\pi(i_\ell)})_{\ell=1}^k \bigr).
\ee

Everything here is exact, and it is rigorous provided we can prove the absolute convergence of the series of cluster terms in \eqref{clexp}.
A sufficient criterion is that, for any $i$,
\be
\label{cond for clexp}
\sum_{B, \supp B \ni i} |\Phi_B(\bsx,\pi)| \e{a|B|} \leq a
\ee
for some constant $a>0$.
See e.g.\ \cite{KP,Uel} for concise statements about cluster expansions.
The sum above involves bonds whose positions are far away.
In order to get such an estimate, one needs to control spatial decay.
It depends on permutations, and there are combinatorial difficulties.

We conclude this subsection by discussing various estimates for the terms above.
Using $1 - \e{-x} \leq x$, we have that
\be
\|\Upsilon\|_\infty = \sup_\omega \Upsilon(\omega) \leq \beta \|U\|_\infty.
\ee
The interesting regime of parameters is $\beta \sim 1/T_{\rm c}^{(0)}$ and $U \to 0$, so $\Upsilon$ is arbitrarily small.
If the potential $U$ is a hard-core with small radius, then $\|\Upsilon\|_\infty = 1$, but $\|\Upsilon\|_p$ is small for $p<\infty$.
We also have that, for any $B$, $\bsx$, and $\pi$,
\be
|\Phi_B(\bsx,\pi)| \leq \|\Upsilon\|_\infty^{|B|}.
\ee
Consider the series \eqref{def Vk} for the potential at order $k$.
The sets $B_1,\dots,B_m$ that contribute to lowest order are such that $\sum_\ell |B_\ell| = k-1$.
It follows that $V^{(k)}$ is of order $\|\Upsilon\|_\infty^{k-1}$.

We can extract the lowest order term.
The expression for $V^{(2)}(x_i,x_{\pi(i)},x_j,x_{\pi(j)})$ involves terms of arbitrary orders.
But we only need to consider $-\Phi_B$ with $B$ containing the single bond $b = \{i,j\}$.
We then obtain the potential $V$ defined in \eqref{def 2 body interaction}.

\subsection{A simpler expression for the interaction}

We now seek to simplify the formula \eqref{def 2 body interaction}.
Namely, we can replace the two integrals over Brownian bridges by a single integral, which will lead to the nicer formula \eqref{LaFormule}.
We have
\bm
\label{on developpe}
\int\dd W_{xy}^{2\beta}(\omega) \int\dd W_{x'y'}^{2\beta}(\omega') \Upsilon(\omega-\omega') = \lim_{n\to\infty} \int_{\Lambda^{2(n-1)}} \dd x_2 \dots \dd x_n \, \dd x_2' \dots \dd x_n' \\
\Bigl[ \prod_{i=1}^n g_{2\beta/n}(x_{i+1}-x_i) g_{2\beta/n}(x_{i+1}'-x_i') \Bigr] \Bigl[ 1 - \exp \Bigl\{ -\frac\beta n \sum_{i=1}^n U(x_i-x_i') \Bigr\} \Bigr]
\end{multline}
with $x_1=x$, $x_{n+1}=y$, $x_1'=x'$, $x_{n+1}'=y'$.
Let us introduce $z_i = x_i-x_i'$.
It is not hard to check that
\be
\label{c'est egal}
g_{2\beta/n}(x_{i+1}-x_i) \, g_{2\beta/n}(x_{i+1}'-x_i') = g_{\beta/n}(x_{i+1} - x_i - \tfrac12 z_{i+1} + \tfrac12 z_i) \, g_{4\beta/n}(z_{i+1}-z_i).
\ee
Substituting into \eqref{on developpe}, we get
\bm
\label{D velo peuh}
\lim_{n\to\infty} \int_{\Lambda^{n-1}} \dd z_2 \dots \dd z_n \Bigl[ \prod_{i=1}^n g_{4\beta/n}(z_{i+1}-z_i) \Bigr] \Bigl[ 1 - \exp \Bigl\{ -\frac\beta n \sum_{i=1}^n U(z_i) \Bigr\} \Bigr] \\
\int_{\Lambda^{n-1}} \dd x_2 \dots \dd x_n \prod_{i=1}^n g_{\beta/n}(x_{i+1} - x_i - \tfrac12 z_{i+1} + \tfrac12 z_i).
\end{multline}
Using Eq.\ \eqref{produit de gaussiennes}, the last line is equal to
\[
g_\beta \bigl( y - x - \tfrac12 (y-y'-x+x') \bigr) = g_\beta \bigl( \tfrac12 (y+y'-x-x') \bigr).
\]
The first line of \eqref{D velo peuh} yields an integral over Brownian paths.
Then \eqref{D velo peuh} is equal to
\be
g_\beta \bigl( \tfrac12 (y+y'-x-x') \bigr) \int \bigl[ 1 - \e{-\frac14 \int_0^{4\beta} U(\omega(s)) \dd s} \bigr] \dd W_{x-x', y-y'}^{4\beta}(\omega).
\ee
Finally, we have the following identity, similar to \eqref{c'est egal}
\be
g_{2\beta}(y-x) g_{2\beta}(y'-x') = g_\beta \bigl( \tfrac12 (y+y'-x-x') \bigr) g_{4\beta}(y-y'-x+x').
\ee
Recall that the two-body interaction defined in \eqref{def 2 body interaction} involves normalized Wiener measures.
Putting normalizations back, we get the following elegant formula for the interaction between jumps $x \mapsto y$ and $x' \mapsto y'$,
\be
\label{LaFormule}
\boxed{V( x, y, x', y') = \int \bigl[ 1 - \e{-\frac14 \int_0^{4\beta} U(\omega(s)) \dd s} \bigr] \dd\widehat W_{x-x', y-y'}^{4\beta}(\omega).}
\ee

It would be useful to obtain a closed form expression in terms of special functions, if it is possible.
When $U$ consists of a hard-core potential of radius $a$, $V( x, y, x', y')$ is equal to the probability that a Brownian bridge, starting at $x-x'$ and ending at $y-y'$, intersects the ball of radius $a$ centered at 0.

\subsection{Effect of interactions on the critical temperature}

The model of spatial permutations should help to clarify the effects of interactions on the critical temperature of Bose-Einstein condensation.

Let $T_{\rm c}^{(a)}$ be the critical temperature for Bose-Einstein condensation as a function of the scattering length $a$ of the interaction potential $U$ between quantum particles.
It is believed that $T_{\rm c}^{(a)}$ behaves in three dimensions as
\be
\label{crittempa}
\frac{T_{\rm c}^{(a)} - T_{\rm c}^{(0)}}{T_{\rm c}^{(0)}} = c \rho^{1/3} a + o(\rho^{1/3} a),
\ee
with $c$ a {\it universal} constant that does not depend on the mass of particles or on the interactions.
The value and even the sign of $c$ has been contested in the physics literature, although a consensus has recently emerged that $c \approx 1.3$.
See \cite{AM,KPS,Kas,NL} and references therein.

The model of spatial permutations is clearly simpler than the Feynman-Kac representation of the Bose gas, and is therefore better suited to Monte-Carlo simulations.
More importantly, we expect that this model, with the interaction \eqref{LaFormule}, is {\it exactly} related to the original quantum boson model, to lowest order in $a$.
Numerical simulations should allow to determine the constant $c$ in the model of permutations with high precision, and with high confidence.
It should be identical to the universal constant of \eqref{crittempa} for the interacting Bose gas.

\section{A simple model of interacting spatial permutations}
\label{sectoymodel}

In this final section, we discuss a simple model of interacting spatial permutations that was introduced in \cite{BU}.
We consider only interactions between permutation jumps of 2-cycles, arguably the most important.
The resulting model is exactly solvable, and it provides a heuristic description for the shift in the critical temperature of the Bose-Einstein condensation.

The approximation consists in replacing the Hamiltonian \eqref{def Ham} by
\be
\tilde H(\bsx,\pi) = \frac1{4\beta} \sum_{i=1}^N |x_i - x_{\pi(i)}|^2 + \sumtwo{1\leq i<j\leq N}{\pi(i)=j, \pi(j)=i} V(x_i,x_j,x_j,x_i).
\ee
The interaction term $V(\cdot)$ is given by \eqref{LaFormule} as before.
From now on, all computations will be exact, at least to lowest order in the scattering length of the original potential $U$.
We consider the three-dimensional case, obviously the most interesting.
A computation shows that
\be
\label{V approx}
V(x,y,y,x) = \frac{2a}{|x-y|} + O(a^2).
\ee
The lowest order term in the right side does not depend on $\beta$, surprisingly.
The expectation of a random variable of permutations is given by \eqref{def expectation},
\be
E_{\Lambda,N}(\theta) = \frac1{Z(\Lambda,N) N!} \sum_{\pi \in \caS_N} \theta(\pi) \int_{\Lambda^N} \dd\bsx \e{-\tilde H(\bsx,\pi)}.
\ee
We now substitute $\tilde H$ with the following simpler Hamiltonian $H^{(\alpha)}$:
\be
\label{Ham toy model}
H^{(\alpha)}(\bsx,\pi) = \frac1{4\beta} \sum_{i=1}^N |x_i - x_{\pi(i)}|^2 + \alpha N_2(\pi),
\ee
with $N_2(\pi)$ denoting the number of 2-cycles in the permutation $\pi$.
The substitution is exact provided that, for any given permutation $\pi$,
\be
\int_{\Lambda^N} \dd\bsx \e{-H(\bsx,\pi)} = \int_{\Lambda^N} \dd\bsx \e{-H^{(\alpha)}(\bsx,\pi)}.
\ee
Isolating the contribution of 2-cycles, this equation reduces to
\be
\int_{\Lambda^2} \dd x_1 \dd x_2 \e{-\frac1{2\beta} |x_1-x_2|^2 - V(x_1,x_2,x_2,x_1)} = \int_{\Lambda^2} \dd x_1 \dd x_2 \e{-\frac1{2\beta} |x_1-x_2|^2 - \alpha}.
\ee
With $V(\cdot)$ in \eqref{V approx}, we find that
\be
\label{relation alpha a}
\alpha = \Bigl( \frac8{\pi\beta} \Bigr)^{1/2} a + O(a^2).
\ee

We now compute the pressure of the model with Hamiltonian $H^{(\alpha)}$.
The grand-canonical partition function is given by
\be
Z'(\beta,\Lambda,\mu) = \sum_{N\geq0} \frac{\e{\beta\mu N}}{N!} \sum_{\pi\in\caS_N} \int_{\Lambda^N} \dd\bsx \e{-H^{(\alpha)}(\bsx,\pi)}.
\ee
It is convenient to work in the Fourier space.
Let us introduce a new partition function,
\be
\label{fpart gc}
Z(\beta,\Lambda,\mu) = \sum_{N\geq0} \frac{\e{\beta\mu N}}{N!} \sum_{k_1,\dots,k_N \in \Lambda^*}
\sum_{\pi \in \caS_N} \e{-\alpha N_2(\pi)} \prod_{i=1}^N \e{-\beta |2\pi k_i|^2} \delta_{k_i, k_{\pi(i)}}.
\ee
Here, $\Lambda^* = \frac1L \bbZ^3$ is the dual lattice.
The thermodynamic pressure is defined by
\be
\label{une pression !}
p^{(\alpha)}(\beta,\mu) = \lim_{V\to\infty} \frac1{\beta V} \log Z(\beta,\Lambda,\mu)
\ee

One can verify that the partition functions $Z$ and $Z'$ differ in two respects only.
First, a normalization is missing, which results in a shift of the chemical potential.
Second, $Z$ has been defined with periodic boundary conditions, unlike $Z'$ (where boundary conditions are neither periodic, nor Dirichlet).
But both partition functions yield the same thermodynamics; precisely,
\be
\lim_{V\to\infty} \frac1{\beta V} \log Z'(\beta,\Lambda,\mu) = p^{(\alpha)} \bigl( \beta, \mu + \tfrac3{2\beta} \log(4\pi\beta) \bigr).
\ee

We now compute $p^{(\alpha)}$.
Introducing occupation numbers, \eqref{fpart gc} becomes
\be
Z(\beta,\Lambda,\mu) = \sum_{(n_k)_{k\in\Lambda^*}} \prod_{k \in \Lambda^*} \biggl[ \e{-\beta (|2\pi k|^2 - \mu) n_k}
\sum_{\pi_k \in \caS_{n_k}} \frac1{n_k!} \e{-\alpha N_2(\pi_k)} \biggr].
\ee
We decomposed the permutation $\pi$ into permutations $(\pi_k)$ for each Fourier mode, and we also used
\be
N_2(\pi) = \sum_{k \in \Lambda^*} N_2(\pi_k).
\ee
Notice that the chemical potential needs to be strictly negative, as in the ideal gas.
We get
\be
\label{palpha}
p^{(\alpha)}(\beta,\mu) = \lim_{V\to\infty} \frac1{\beta V} \sum_{k\in\Lambda^*} \log \biggl[ \sum_{n\geq0} \e{-\beta (|2\pi k|^2 - \mu) n}
\sum_{\pi \in \caS_n} \frac1{n!} \e{-\alpha N_2(\pi)} \biggr].
\ee
Let us compute the bracket above.
For given $\pi \in \caS_n$, let $r_j$ denote the number of cycles of length $j$.
Then $\sum_j j r_j = n$, and the number of permutations for given $(r_j)$ is equal to
\[
n! \Big/ \prod_{j\geq1} j^{r_j} r_j!.
\]
The bracket in \eqref{palpha} is equal to
\[
\begin{split}
\sum_{n\geq0} \frac1{n!} & \sumtwo{r_1,r_2,\dots \geq 0}{\sum_j j r_j = n} \frac{n!}{\prod_{j\geq1} j^{r_j} r_j!} \e{-\beta (|2\pi k|^2-\mu) \sum_j j r_j} \e{-\alpha r_2} \\
&= \sum_{r_1,r_3,r_4,\dots\geq0} \prod_{j=1,3,4,\dots} \frac1{r_j!} \bigl[ \tfrac1j \e{-j\beta (|2\pi k|^2-\mu)} \bigr]^{r_j}
\sum_{r_2\geq0} \frac1{r_2!} \bigr[ \tfrac12 \e{-2\beta (|2\pi k|^2-\mu) - \alpha} \bigr]^{r_2} \\
&= \exp \Bigl\{ \sum_{j=1,3,4,\dots} \tfrac1j \e{-j\beta (|2\pi k|^2-\mu)} + \tfrac12 \e{-2\beta (|2\pi k|^2-\mu) - \alpha} \Bigr\} \\
&= \exp \Bigl\{ -\log (1 - \e{-\beta (|2\pi k|^2-\mu)}) - \tfrac12 \e{-2\beta (|2\pi k|^2-\mu)} (1-\e{-\alpha}) \Bigr\}.
\end{split}
\]
This can be inserted into \eqref{palpha}.
In the limit $V\to\infty$ the expression converges to a Riemann integral.
If $\alpha=0$, we get the pressure of the ideal gas
\be
\label{idealpressure}
p^{(0)}(\beta,\mu) = - \frac1\beta \int_{\bbR^3} \log \bigl( 1 - \e{-\beta (|2\pi k|^2-\mu)} \bigr) \dd k,
\ee
as expected.
And if $\alpha \neq 0$, we get
\be
p^{(\alpha)}(\beta,\mu) = p^{(0)}(\beta,\mu) - \frac{\e{2\beta\mu}}{2^{11/2} \pi^{3/2} \beta^{5/2}} (1-\e{-\alpha}).
\ee

\bfig
\epsfxsize=120mm \centerline{\epsffile{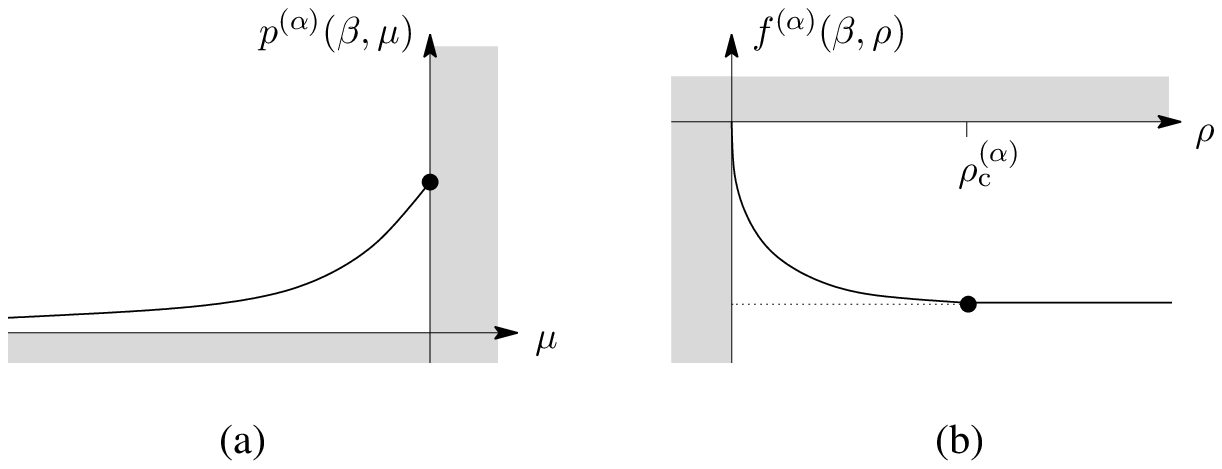}}
\caption{The pressure and the free energy of the simple interacting model in three dimensions.}
\label{figpf}
\efig

The pressure $p^{(\alpha)}$ is plotted in Fig.\ \ref{figpf} (a) as a function of $\mu$.
One can consider other thermodynamic potentials as well.
Recall that the free energy $f^{(\alpha)}$ is function of the (inverse) temperature and of the density, and it is related to the pressure by a Legendre transform:
\be
f^{(\alpha)}(\beta,\rho) = \sup_\mu \bigl[ \rho\mu - p^{(\alpha)}(\beta,\mu) \bigr].
\ee
One then obtains the graph depicted in Fig.\ \ref{figpf} (b).
It is strictly decreasing up to the {\it critical density} $\rho_{\rm c}^{(\alpha)} = \rho_{\rm c}^{(\alpha)}(\beta)$, and it is constant afterwards.
The critical density is equal to the derivative of $p^{(\alpha)}$ with respect to $\mu$ at $0-$.
We have
\be
\label{critdensalpha}
\rho_{\rm c}^{(\alpha)} = \rho_{\rm c}^{(0)} - \frac1{2^{9/2} \pi^{3/2} \beta^{3/2}} (1-\e{-\alpha}).
\ee
The first term of the right side, $\rho_{\rm c}^{(0)}$, is equal to the critical density of the ideal gas, Eq.\ \eqref{critdens}.
The second term is the correction due to our simple interaction.

We see that $\rho_{\rm c}^{(\alpha)}$ is smaller than $\rho_{\rm c}^{(0)}$ --- interactions favour Bose-Einstein condensation.
This observation is in line with physicists' expectations.
The heuristics is particularly simple in this model:
2-cycles are penalized and this favours all other cycles, including infinite cycles.
The latter occur therefore at a lower density.
While elementary, this heuristics may well be correct.

Let us now estimate the change in the critical temperature.
Using \eqref{relation alpha a} with $\beta = 1/T_{\rm c}^{(0)}$, we find that, to lowest order,
\be
\frac{T_{\rm c}^{(a)} - T_{\rm c}^{(0)}}{T_{\rm c}^{(0)}} = \tfrac1{3\sqrt2} \zeta(\tfrac32)^{-1} \alpha = \tilde c \, \rho^{1/3} a
\ee
with $\tilde c = 0.37$.
This formula can be compared to \eqref{crittempa}.
If we believe the value $c = 1.3$ found numerically, then 2-cycle interactions account only for a fraction of the effect of all interactions.
One could also take into account the interactions within 3-cycles and longer cycles;
the constant $\tilde c$ would increase a bit.

One would expect infinite cycles to occur for all densities larger than the critical density \eqref{critdensalpha}.
More precisely, Theorem \ref{thminfinitecycles} should remain valid for $\alpha>0$, replacing $\rho_{\rm c}$ by $\rho_{\rm c}^{(\alpha)}$.
But only a weaker claim has been proved so far.

\begin{theorem}
\label{thmalpha}
For any $b<1$,
\[
\lim_{V\to\infty} E_{\Lambda,\rho V} (\bsvarrho_{V^b, \rho V}) \geq \rho - \frac4{(1+\e{-\alpha})^2} \rho_{\rm c}^{(0)}.
\]
\end{theorem}

Theorem \ref{thmalpha} guarantees the existence of macroscopic cycles for large enough densities.
The proof can be found in \cite{BU}.

\bigskip
{\bf Acknowledgments.}
I would like to thank V.\ Betz for useful comments and discussions.
I am also grateful to Ingrid Belti\c ta, Gheorghe Nenciu, and Radu Purice for organising the conference QMath 10 in Moeciu, Romania, in September 2007.
And to V.\ Zagrebnov for organising the special session ``Condensed Matter \& Open Systems'' where the present subject was presented.
This work is supported in part by the grant DMS-0601075 of the US National Science Foundation.


\begin{thebibliography}{99}

\bibitem{AM}
P.\ Arnold, G.\ Moore,
{\it BEC Transition Temperature of a Dilute Homogeneous Imperfect Bose Gas},
Phys.\ Rev.\ Lett.\ 87, 120401 (2001)

\bibitem{BU}
V.\ Betz, D.\ Ueltschi,
{\it Spatial random permutations and infinite cycles},
preprint, arxiv:0711.1188 (2007)

\bibitem{Fey}
R.\ P.\ Feynman,
{\it Atomic theory of the $\lambda$ transition in Helium},
Phys.\ Rev.\ 91, 1291--1301 (1953)

\bibitem{GRU}
D.\ Gandolfo, J.\ Ruiz, D.\ Ueltschi,
{\it On a model of random cycles},
J.\ Stat.\ Phys.\ 129, 663--676 (2007)

\bibitem{Gin}
J.\ Ginibre,
{\it Some applications of functional integration in statistical mechanics},
in ``M\'ecanique statistique et th\'eorie quantique des champs'', Les Houches 1970, C.\ DeWitt and R.\
Stora eds, 327--427 (1971)

\bibitem{HS}
C.\ Hainzl, R.\ Seiringer,
{\it General decomposition of radial functions on $\bbR^n$ and applications to $N$-body quantum systems},
Lett.\ Math.\ Phys.\ 61, 75--84 (2002)

\bibitem{KPS}
V.\ A.\ Kashurnikov, N.\ V.\ Prokof'ev, B.\ V.\ Svistunov,
{\it Critical temperature shift in weakly interacting Bose gas},
Phys.\ Rev.\ Lett.\ 87, 120402 (2001)

\bibitem{Kas}
B.\ Kastening,
{\it Bose-Einstein condensation temperature of a homogenous weakly interacting Bose gas in variational perturbation theory through seven loops},
Phys.\ Rev.\ A 69, 043613 (2004)

\bibitem{KP}
R.\ Koteck\'y, D.\ Preiss,
{\it Cluster expansion for abstract polymer models},
Comm.\ Math.\ Phys.\ 103, 491--498 (1986)

\bibitem{NL}
K.\ Nho, D.\ P.\ Landau,
{\it Bose-Einstein condensation temperature of a homogeneous weakly interacting Bose gas: Path integral Monte Carlo study},
Phys.\ Rev.\ A 70, 053614 (2004)

\bibitem{PC}
E.\ L.\ Pollock, D.\ M.\ Ceperley,
{\it Path-integral computation of superfluid densities},
Phys.\ Rev.\ B 36, 8343--8352 (1987)

\bibitem{Suto1}
A.\ S\"ut\H o, {\it Percolation transition in the Bose gas}, J.\
Phys.\ A 26, 4689--4710 (1993)

\bibitem{Suto2}
A.\ S\"ut\H o,
{\it Percolation transition in the Bose gas II},
J.\ Phys.\ A 35, 6995--7002 (2002)

\bibitem{Uel}
D.\ Ueltschi,
{\it Cluster expansions and correlation functions},
Moscow Math.\ J.\ 4, 511--522 (2004); math-ph/0304003

\bibitem{Uel1}
D.\ Ueltschi,
{\it Feynman cycles in the Bose gas},
J.\ Math.\ Phys.\ 47, 123302 (2006)

\end{thebibliography}
\end{document}